\documentclass[structabstract]{aa}  
\usepackage{graphicx}
\usepackage{txfonts}
\begin{document}
   \title{Day and night side core cooling of a strongly irradiated giant 
planet}


   \author{J. Budaj \inst{1}
\and I. Hubeny \inst{2,1}
\and A. Burrows \inst{3}
          }

   \institute{Astronomical Institute, Slovak Academy of Sciences,
              05960 Tatransk\'{a} Lomnica, Slovak Republic,
              \email{budaj@ta3.sk}
\and
Steward Observatory and Dept. of Astronomy, University of Arizona, Tucson, USA,
\email{hubeny@as.arizona.edu}
\and
Dept. of Astrophysical Sciences, Princeton, USA,
\email{burrows@astro.princeton.edu}
             }

   \date{Received Aug 30, 2011; accepted Nov 9, 2011}

 
  \abstract
   {
The internal heat loss, or cooling, of a planet determines its structure
and evolution.
}
   {
We study the effects of irradiation,
metallicity of the atmosphere, heat redistribution, stratospheres,
and the depth where the heat redistribution takes place on 
the atmospheric structure, the core entropy, and subsequently on 
the cooling of the interior of the planet.
}
   {
We address in a consistent fashion the coupling between
the day and the night sides of a planet by means of model atmosphere 
calculations with heat redistribution. We assume that strong convection 
leads to the same entropy on the day and night sides and that gravity is
the same on both hemispheres.
}
   {
We argue that the core cooling rates from the two
hemispheres of a strongly irradiated planet may not be the same and
that the difference depends on several important parameters.
If the day-night heat redistribution is very efficient or if it takes place
at the large optical depth, then the day-side and the night-side cooling may
be comparable. However, if the day-night heat transport is not
efficient or if it takes place at a shallow optical depth then there can be
a large difference between the day- and the night-side cooling and
the night side will cool more efficiently.
If stellar irradiation becomes stronger, e.g. owing to stellar evolution
or migration, cooling from both the day and the night sides is reduced.
Enhanced metallicity of the atmosphere would act as an added ``blanket" 
and reduces both the day- and the night-side cooling.
However, a stratosphere on the planetary day side can enhance
day-side cooling since its opacity acts as a ``sunshade" that screens
the stellar irradiation.
These effects may also influence the well-known gravity darkening and 
bolometric albedo effects in interacting binaries, especially for strongly 
irradiated cold components.
}
{}

\maketitle
%

\section{Introduction}

During its evolution, a planet derives heat from both its gravitational energy 
and other possible extra heat sources. In particular, tidal heating 
(Jackson, Greenberg, \& Barnes \cite{jgb08};
Liu, Burrows, \& Ibgui \cite{lbi08}) or ohmic heating 
(Batygin, Stevenson, \& Bodenheimer \cite{bsb11};
Perna, Menou \&  Rauscher \cite{pmr10}) may be quite important 
for some close-in extrasolar planets.
Also, a small fraction of received stellar irradiation may be
transported to deeper layers (Guillot \& Showman \cite{gs02}).
This energy is radiated from the planet via its atmosphere, which 
acts as a valve that regulates cooling.
A close-in planet is subject to strong stellar irradiation
that is many orders of magnitude stronger than the internal heat
loss of the planet. Subsequently, the structure of the atmosphere 
and the emergent spectrum are governed by the stellar irradiation.

Incoming stellar radiation does not penetrate very deeply
into the interior of a planet and is effectively reradiated.
Consequently, for any given assumed chemical composition, the structure of
the planet itself is governed by its internal heat content, together
with possible, additional energy sources.
This has direct observational consequences. As a planet evolves, 
its interior cools and its radius shrinks, even if the planet is exposed 
to enhanced stellar irradiation due to stellar evolution or the planet's migration.
Planet cooling, however, can be significantly affected by stellar 
irradiation, which can keep the planet inflated for a longer
period of time (Guillot et al. \cite{gbh96}). 

Throughout this paper, we concentrate solely on close-in planets for
which the spin and orbital periods are forced to be equal.
Consequently, the planet surface is naturally 
divided between a day side that permanently faces the star and a night side 
that experiences no direct stellar irradiation.
The irradiation energy is scattered, absorbed, and reradiated on the day
side of the planet. Moreover, a fraction of the total incoming stellar 
energy is carried to the night side of the planet, where it can be 
reradiated. This complex radiative and hydrodynamical mechanism determines
the properties of the atmosphere, and, thus, also planet cooling.
Generally, the planet core may cool at a very different rate on the day 
and on the night sides. The magnitude of cooling depends mainly on 
the magnitude of the stellar irradiation, the amount of the day-night heat redistribution, 
the depth where this day-night heat redistribution takes place, and 
the atmospheric structure and opacities.
Previous evolutionary studies (Laughlin et al. \cite{lwv05};
Burrows et al. \cite{bhb07}; Fortney, Marley, \& Barnes \cite{fmb07};
Liu, Burrows, \& Ibgui \cite{lbi08}) have 
implicitly assumed full heat redistribution, i.e. that the night-side
atmosphere is identical with the one on the day side and that the core heat 
loss from the night side is equal to that from the day side. 
The effect of the irradiation and the day-night heat redistribution
was achieved by simply reducing the irradiating stellar flux by    
a factor of two prior to its encounter with the atmosphere of the planet
(e.g., Sudarsky, Burrows, \& Hubeny \cite{sbh03}).
Recently, Guillot (\cite{guillot10}) and Heng et al.(\cite{hh11}) 
have constructed analytical gray irradiated atmosphere models to study 
planet cooling. 

Irradiation effects have also been studied in the field of interacting binaries.
Irradiated models of the atmosphere have been constructed by
Rucinski (\cite{rucinski70}, \cite{rucinski73}).
Rucinski (\cite{rucinski69}) and Vaz \& Norlund (\cite{vn85})
estimated the bolometric albedo for irradiated stars having deep 
adiabatic convective envelopes, assuming constant entropy. 

The aim of this paper is to assess, in a systematic way,
differences between the day and the night side core cooling rates, and
to study the effects of the strength of stellar irradiation,
atmospheric metallicity, the efficiency of day-night heat
redistribution, and the depth where this redistribution takes place.
We use a state-of-the-art model atmosphere code, where the day-night heat
redistribution is treated in a parametric way.
We illustrate the effects on the prototype of a close-in
extrasolar planet: HD\,209458b. In Sec.\ref{s2}, we describe  
the calculations and in Sections \ref{s3}-\ref{s6} we study    
the effects of the various parameters that are important for     
the day- and night-side core cooling rates.

\section{Modeling procedure}
\label{s2}

As is well known from planetary evolution models, the interior of the planet 
is fully convective (e.g., Hubbard, Burrows \& Lunine \cite{hbl02}),
and that the convection is very nearly adiabatic.
We make the realistic assumption that the entropy in the convection zone
is constant everywhere in the planet. Specifically, the entropies at 
the base of the convection zone at the day and night sides are equal.
Moreover, we assume that the planet is spherical, so that the gravitational
accelerations on the day and the night sides are equal. 
This is also a reasonable assumption for our purpose.
Budaj (\cite{budaj11}) calculated shapes and variations in the effective
gravity over the surface of all currently known transiting exoplanets.
He found that largest departures from the spherical symmetry are expected 
for WASP-12b and WASP-19b, of about 12-15\%, respectively which translates 
into the variations of the gravity over the surface of about
25-32\%. The gravity difference between the substellar and antistellar
points were less than 2\%.
For convenience, we assume that the thickness of the atmosphere is small 
compared to the planetary radius, so that the gravity is a constant.

The entropy of the planetary core is thus determined by the
entropy at the base of the convection zone, which in turn is
determined by the magnitude of stellar irradiation, radiation transport
in the atmosphere, and other details of the atmospheric physics.
Therefore, sophisticated evolutionary models have to take into
account proper boundary conditions, in particular the values of the
entropy and the radiation cooling, using detailed atmospheric models, 
as was done e.g. in
Burrows, Sudarsky, \& Hubbard (\cite{bsh03}),
Burrows et al. (\cite{bhb07}), Fortney, Marley, \& Barnes (\cite{fmb07}), 
and Liu, Burrows, \& Ibgui (\cite{lbi08}).

The amount of radiation cooling from the day and the night sides
is calculated in the following way: From the definition of the effective
temperature, the total energy flux incoming at a unit surface area 
at the lower base of an atmosphere is given by
$F= \sigma T_{\rm eff}^4$.
Since no energy is being lost in the atmosphere, this quantity also 
represents the total radiation flux escaping from the unit surface at 
the upper boundary of the atmosphere. Therefore, it provides a quantitative 
measure of the radiation heat loss through the planet's atmosphere, and,
thus, from the whole planet. 
For usual atmospheric models, the effective temperature, 
together with surface gravity (and overall chemical composition) are taken 
as fundamental parameters, while from the point of view of interior and 
evolutionary models, the primary parameter is the core entropy.

We divide the whole planetary surface into two parts, and treat the
overall planetary atmosphere as composed of two distinct atmospheres --
an averaged day-side atmosphere and an averaged night-side atmosphere.
The effective temperature on the day side is called $T_{d}$, and on the
night side $T_{n}$. To avoid confusion with other possible meanings of 
the term ``effective temperature'' used for instance in planetary science, 
we will call $T_{d}$ and $T_{n}$ ``intrinsic effective temperatures''. 
Obviously, these effective temperatures should not be confused with
brightness temperatures or
an actual atmospheric temperature, which is a function of depth in 
the atmosphere and strongly depends on the irradiation.

As mentioned above, to compute the total heat loss consistently, 
the model atmospheres for the day and the night side must have the same 
entropy at the convective base.  To this end, 
we calculate a grid of models with/without the irradiation
corresponding to the day/night side of the planet for a range of   
day- and night-side intrinsic effective temperatures 
and surface gravities ($\log g$). Each model has a certain entropy
in the convection zone. In the next step we match the entropy and
gravity of the day and night sides, which results in different
effective temperatures ($T_{d}\neq T_{n}$) for the day and night sides.
Notice that we can obtain a unique match because entropy is a 
monotonic function of the intrinsic effective temperature.
Since $T_{d}$ and $T_{n}$ represent the total radiation flux on the day
and night sides, they represent the day and the night side internal
heat loss, or, equivalently, the cooling of the interior. 
The total internal heat loss (cooling),
$L_{\rm cool}$, from the planet is then given by
\begin{equation}
L_{\rm cool}=  4 \pi R_{p}^{2} \sigma T_{\rm eff}^{4} =
2 \pi R_{p}^{2} \sigma (T_{d}^{4}+T_{n}^{4}) ,
\end{equation}
where $R_{p}$ is the radius of the planet, $\sigma$ is
the Stefan-Boltzmann constant, and $T_{\rm eff}$ is the composite
intrinsic effective temperature.

The individual model atmospheres are computed using the code 
{\sc CoolTlusty} designed to model atmospheres of irradiated giant planets   
and brown dwarfs. This code is a version of the stellar atmosphere code
{\sc Tlusty} (Hubeny \cite{hubeny88}; Hubeny \& Lanz \cite{hl95});
modifications for {\sc CoolTlusty} are described in 
Hubeny, Burrows \& Sudarsky (\cite{hbs03}) and Sudarsky, Burrows \& Hubeny
(\cite{sbh03}).
The code solves the hydrostatic and the radiative+convective equilibrium 
equations, assumes LTE, and can take into account clouds and departures 
from local chemical equilibrium (Hubeny \& Burrows \cite{hb07}).
The computed models represent separately averaged-day and averaged-night 
side atmospheres. The upper boundary is set to the pressure
$10^{-5}$ bars and extends deeply into the convection zone.
At a specified depth range, we allow for an energy removal on 
the day side, and an energy deposition on the night side of the planet,
using the procedure described in Burrows, Budaj \& Hubeny (\cite{bbh08}).
The amount of the heat redistribution between the day and night sides 
is parametrized by the $P_{n}$ parameter, which is defined as the fraction 
of the incoming stellar irradiation that is transferred from the day side
to the night side and radiated from there
(Burrows, Sudarsky \& Hubeny \cite{bsh06}).

We consider here the well-known planet HD 209458b.
If not stated otherwise, we assume solar chemical composition   
of the planetary atmosphere, energy removal/deposition at 0.03-0.3 bar, 
and $P_{n}=0.3$. TiO and VO opacity is not considered.
Opacities are taken from Sharp \& Burrows (\cite{sb07}),
assuming chemical-equilibrium compositions with rainout but  
no cloud opacity. Kurucz (\cite{kurucz93}) spectrum of the parent star
HD 209458 is used as the source of irradiation.
Parameters of the star and planet are taken from
Henry et al. (\cite{hmb00}), Charbonneau et al. (\cite{cbl00}), and 
Knutson et al. (\cite{kcn07}).
We assume $T_{\rm eff}=6000$ K, $R=1.13\ R_{\odot}$ for the star,
and a semi-major axis of the planet's orbit $a=0.045$~AU.
The models were calculated for a grid of effective temperatures
ranging from 50 K up to 300 K and gravities from 2.4 up to 3.6 (cgs).

\begin{figure}
\centerline{
\includegraphics[angle=-90,width=8.5cm]{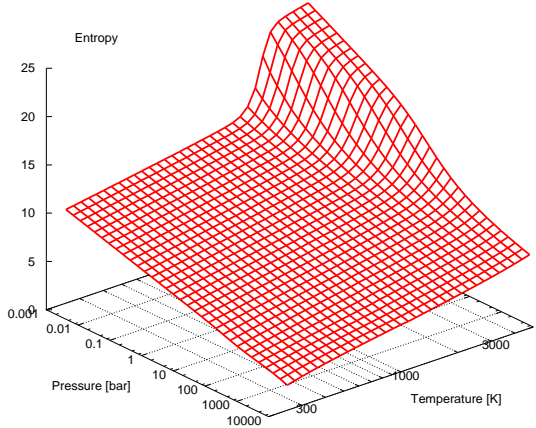}
}
\caption{
{\bf Entropy} (per baryon and divided by $k$) as a function of pressure
and temperature. Entropy decreases with pressure and increases with 
temperature.
}
\label{ent}
\end{figure}

\begin{figure}
\centerline{
\includegraphics[angle=-90,width=8.5cm]{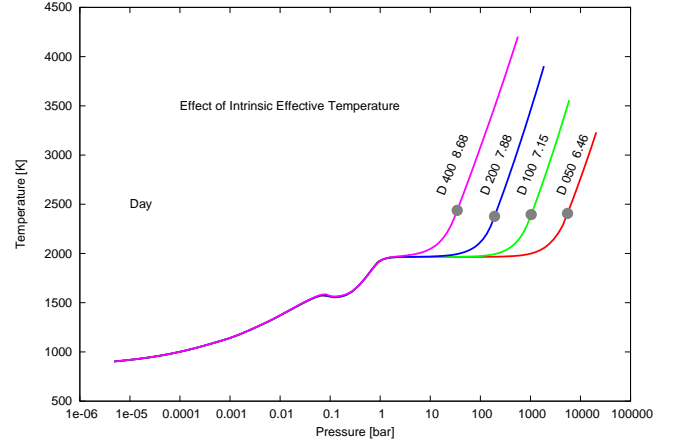}
}
\caption{
{\bf The effect of the intrinsic effective temperature on
the $T$-$P$ profile of the atmosphere on the day side.} 
D - stands for day, numbers give the effective temperature in K
and the core entropy, respectively.
Circles mark the top of the convection zone.
Notice that for higher effective temperatures the convection zone starts at
lower pressures (densities) but at about the same temperatures.
Consequently, given the relation from Fig.\ref{ent}, models with 
higher intrinsic effective temperature have higher core entropy.
}
\label{tefftp}
\end{figure}

We plot in Fig. \ref{ent} the entropy (per baryon, divided by 
the Boltzmann constant $k$) as a function of temperature and pressure
(Saumon, Chabrier, \& Van Horn \cite{sc95}).
The crucial point is that the entropy increases with temperature 
and decreases with pressure. 
In Fig. \ref{tefftp} we plot the temperature - pressure ($T$-$P$) 
profiles as a function of the intrinsic effective temperature all 
else being equal. 
We assumed $P_{n}=0.5$ and irradiated day side of the planet. 
This temperature affects mainly the deeper layers and shifts 
the location of the top of the convection zone.
The higher the intrinsic effective temperature the lower  
the pressure at the convection boundary and, given the relation from
Figure \ref{ent}, the higher the core entropy.
This monotonically changes the entropy in the convection zone. 
It means that we can use this intrinsic effective temperature to adjust 
the entropy in the convection zone on the day and night sides.

\section{The effect of the stellar irradiation on day-night 
side cooling}
\label{s3}

\begin{figure*}[t]
\centerline{
\includegraphics[angle=-90,width=8.5cm]{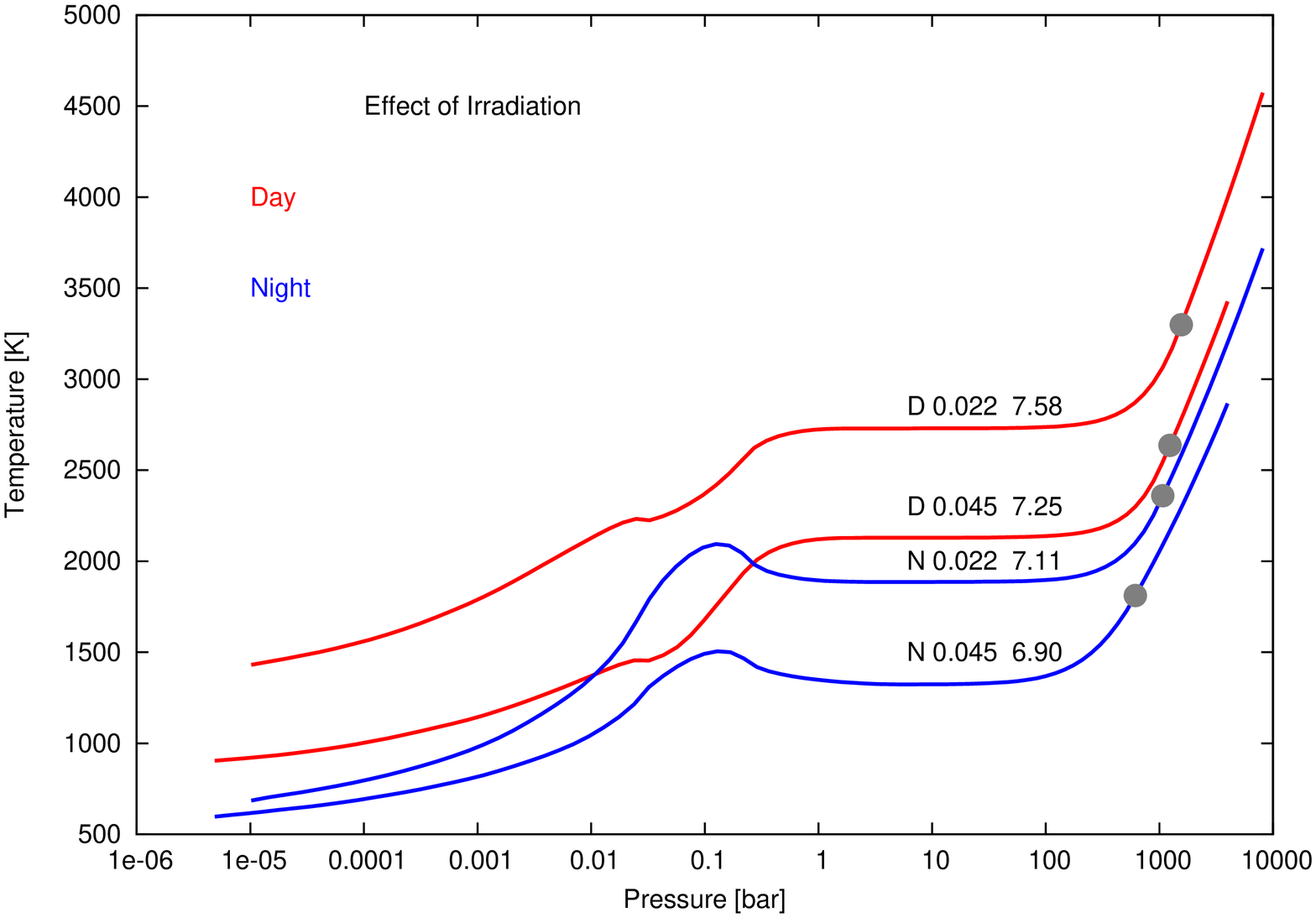}
\includegraphics[angle=-90,width=8.5cm]{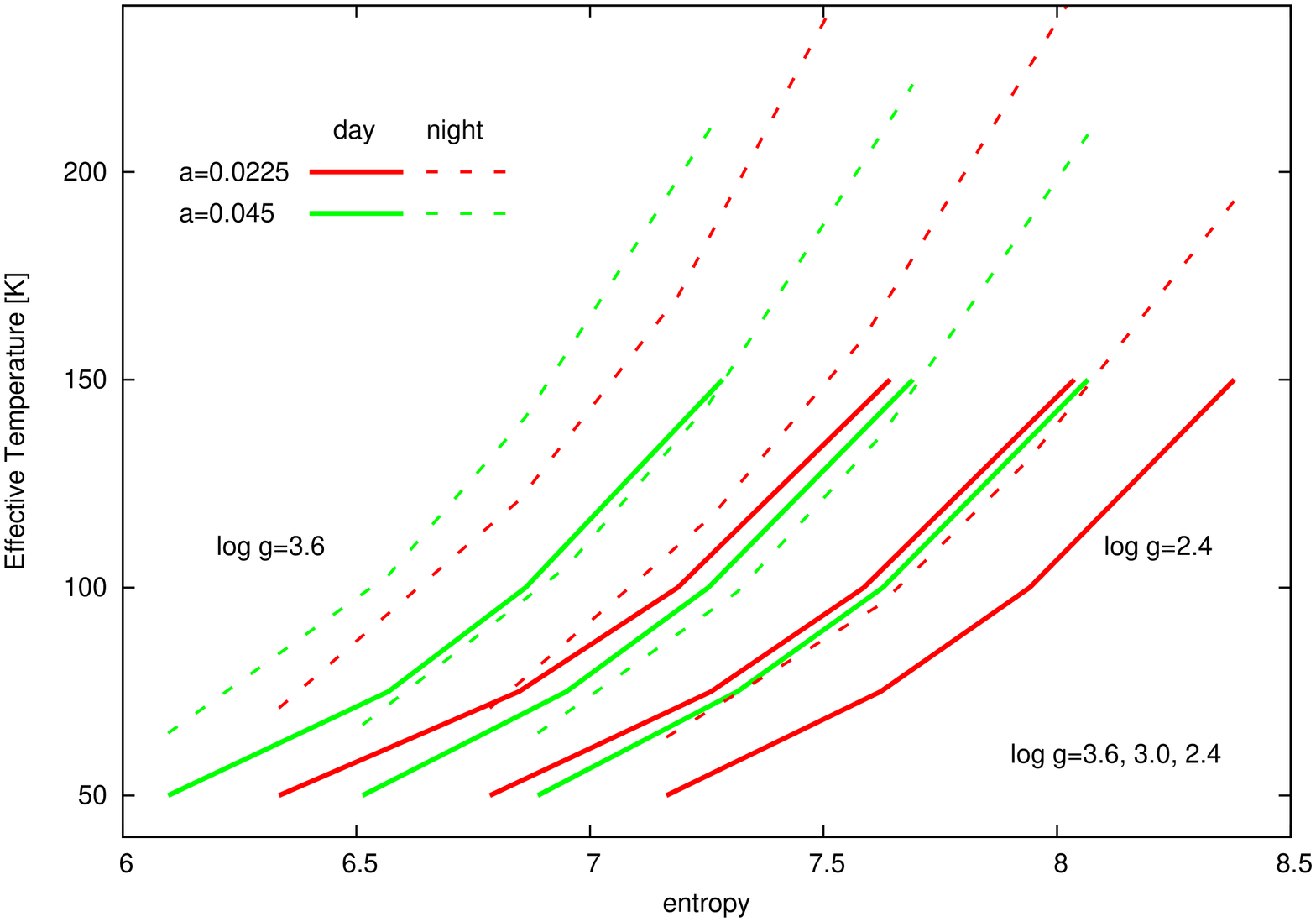}
}
\caption{
{\bf Left: The effect of the planet-star distance (a magnitude of irradiation) on 
the $T$-$P$ profile of the atmosphere.} 
Red lines represent the day side, blue lines represent the night side. 
Circles mark the top of the convection zone.
D - stands for day, N - stands for night, numbers give the planet-star
distance in AU and the core entropy, respectively.
Notice that entropy on both day and night sides increases with increasing
irradiation (shorter distance).
{\bf Right: The effect of the planet-star distance or
irradiation on the day-night side cooling of the planet.}
The cooling is expressed as the intrinsic effective temperature in $K$
as a function of the core entropy for several surface gravities.
The day-side is solid, and night-side is dashed. The models are
calculated for two planet-star distances $0.045$ (green) and $0.0225$~AU
(red).
The cooling from the day side decreases with 
the stellar irradiation (shorter distance). The cooling from
the night side behaves in a similar way.
Consequently, the total heat loss is lower at a higher irradiation.
}
\label{dd1tp}
\end{figure*}

Planets are generally exposed to very different intensities of 
stellar irradiation. Solar-type stars on the main sequence
increase in luminosity in the course of their evolution. 
Planets migrate as a result of their interaction with the 
protoplanetary disk or with their host stars.
Therefore, we first explore the effects of 
irradiation on the day- and night-side cooling. 

Stellar irradiation considerably modifies the day side 
$T$-$P$ profile of the atmosphere.
It flattens the temperature gradient and creates a noticeable 
plateau (Sudarsky, Burrows \& Hubeny \cite{sbh03}; 
Hubeny, Burrows \& Sudarsky \cite{hbs03}). 
Because of day to night side heat transport, this plateau has a signature
on the night side $T$-$P$ profile  as well 
(Burrows, Budaj \&  Hubeny \cite{bbh08}). To study the effect of 
irradiation on the core heat loss we calculate a set of day- and 
night-side atmosphere models at two different planet--star distances,
one corresponding to the nominal distance $a=0.045$ AU, and the other 
two times closer: $a=0.0225$ AU, which approximately
corresponds to a 4 fold increase in stellar irradiation. 
We set $P_{n}=0.3$; and assume that the incoming irradiation energy is 
removed at the depth region between 0.03 and 0.3 bar on the day side, 
and is deposited on the night side at the same range of pressures.

$T$-$P$ profiles for a model with $T_{\rm eff}=100$ K and $\log_{10} g=3$
are shown in Figure \ref{dd1tp} (left).
 Notice that the position of the top of the convection zone varies;
this is exactly the location of the point that determines the entropy in 
the convection zone. The location of this point closely correlates with 
the temperature of the plateau. 
On the day side, the temperature plateau and the top of the convection 
zone are hotter for higher irradiation. Given the dependence from 
Figure \ref{ent} the core entropy is also higher for higher irradiation 
(closer distance). 
This means that one has to 
lower the day side internal effective temperature (keeping $g$ constant) 
to obtain the same entropy as on the night side.
Consequently, this means that we obtain less cooling from the day side.
The night side behaves similarly, since its
$T$-$P$ profiles reflect that of the day side.

\begin{figure*}
\centerline{
\includegraphics[angle=-90,width=10.cm]{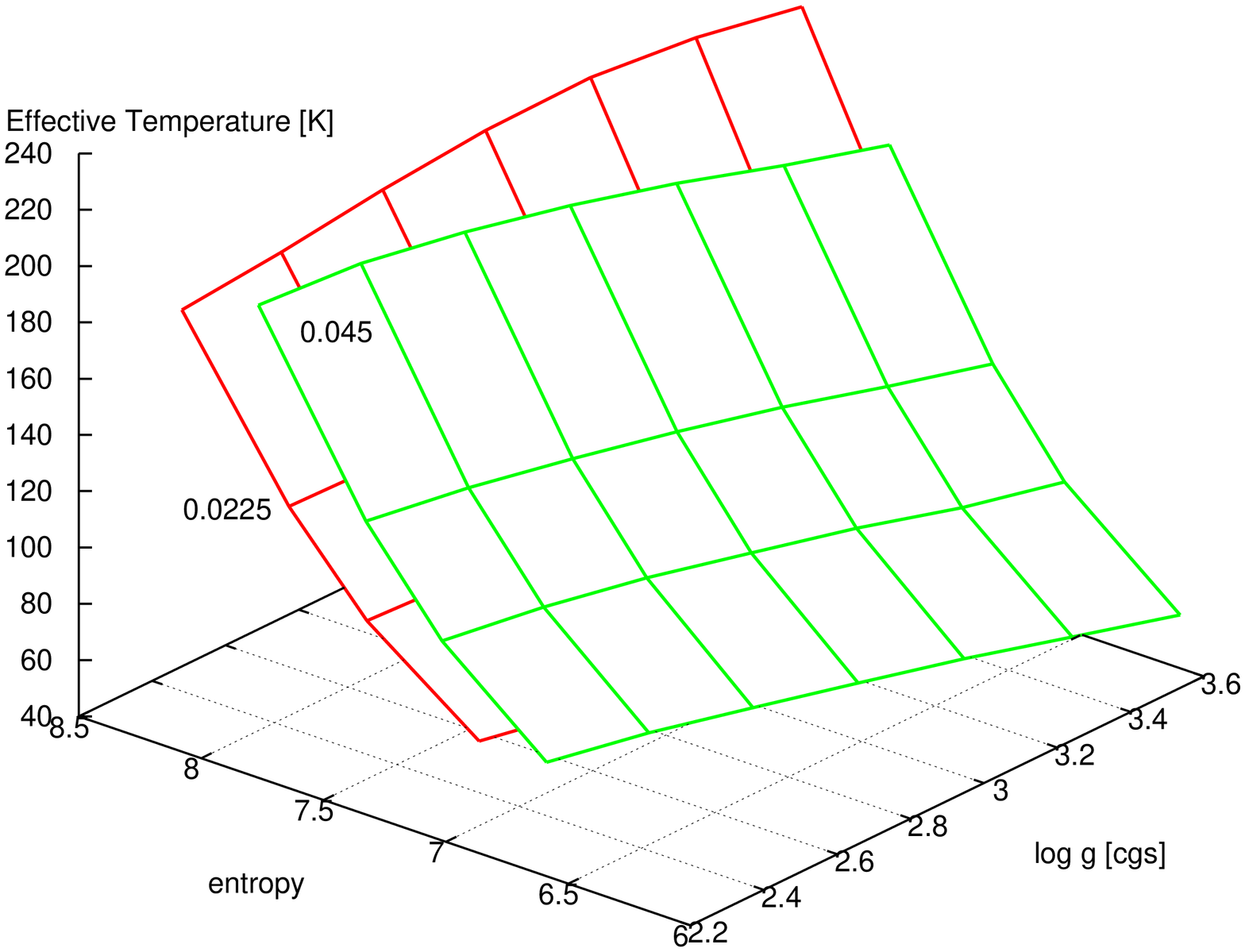}
\hspace{-10mm}
\includegraphics[angle=-90,width=10.cm]{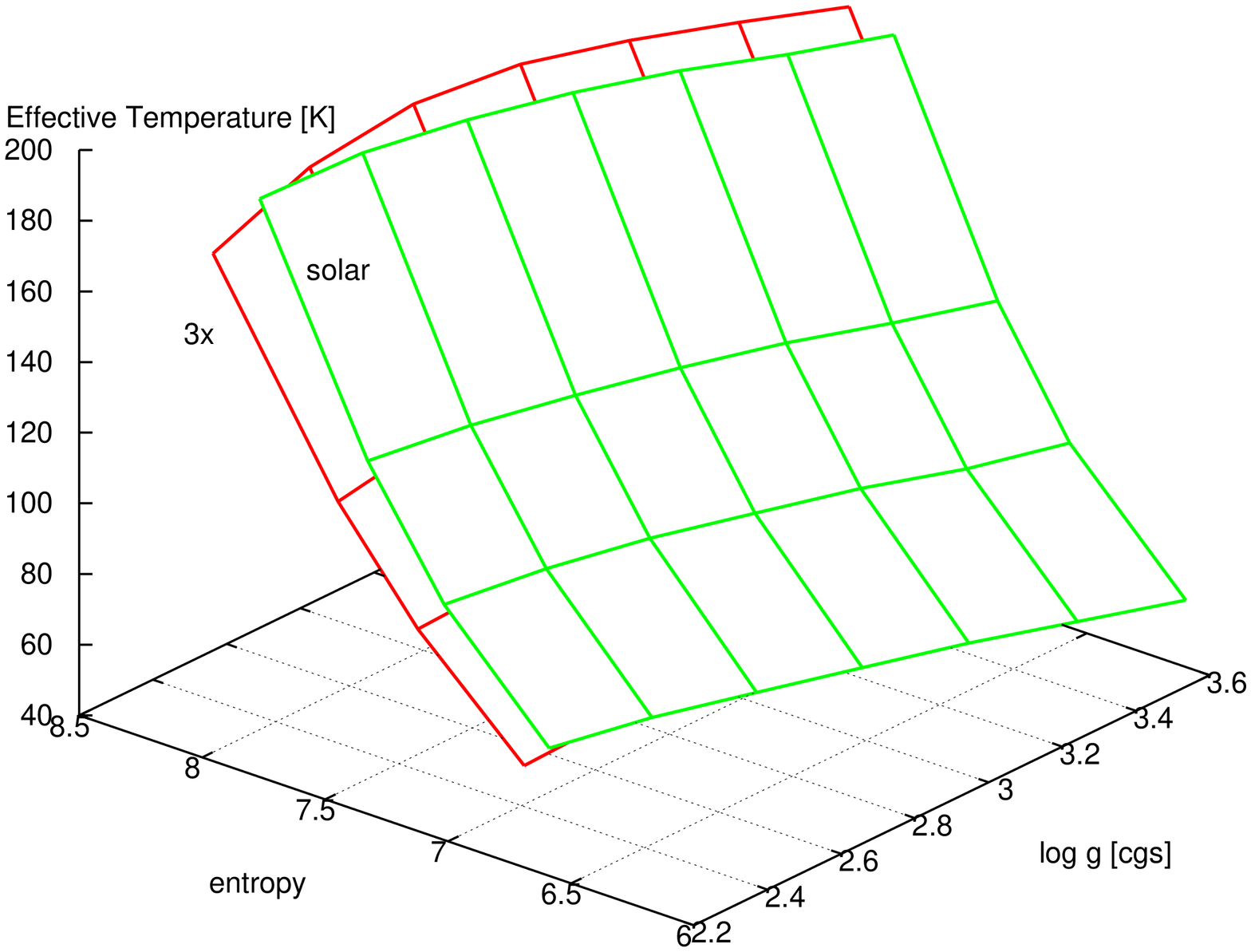}
}
\caption{
The combined night and day side cooling
as a function of the surface gravity and core entropy.
It is expressed as the intrinsic effective temperature
$T_{\rm eff}=(T_{d}^{4}/2+T_{n}^{4}/2)^{0.25}$ in $K$.
{\bf Left: The effect of the irradiation.} The models are
calculated for two different values of the planet-star distance:
0.045, 0.0225 AU.
The total core heat loss decreases with increasing irradiation 
(smaller distance). 
{\bf Right: The effect of the atmospheric metallicity.} The models are
calculated for two different values of the metallicity: solar and 3x solar.
Total heat loss decreases with increasing the metallicity.
}
\label{dd2}
\end{figure*}

Figure \ref{dd1tp} (right) illustrates the intrinsic effective
temperature, and thus core cooling, as 
a function of entropy for several gravities. 
Notice that both the day- and the night-side core 
cooling rates decrease with increasing irradiation. Night side cooling 
is more efficient. This means that for a higher irradiation flux 
the total heat loss from both the day and the night side is lower
(Figure \ref{dd2} left).
This somewhat counterintuitive behavior is understood by noticing
that the irradiation heats the atmosphere, flattens the temperature 
gradient, and thus creates a ``barrier'' that inhibits the heat loss 
from the interior (Burrows et al. \cite{burrows00}).

In conclusion, if the star is brighter or the planet-star 
separation is lower, the total heat loss from a planet is reduced.

\section{The effect of atmospheric metallicity}
\label{s3a}

\begin{figure*}
\centerline{
\includegraphics[angle=-90,width=8.5cm]{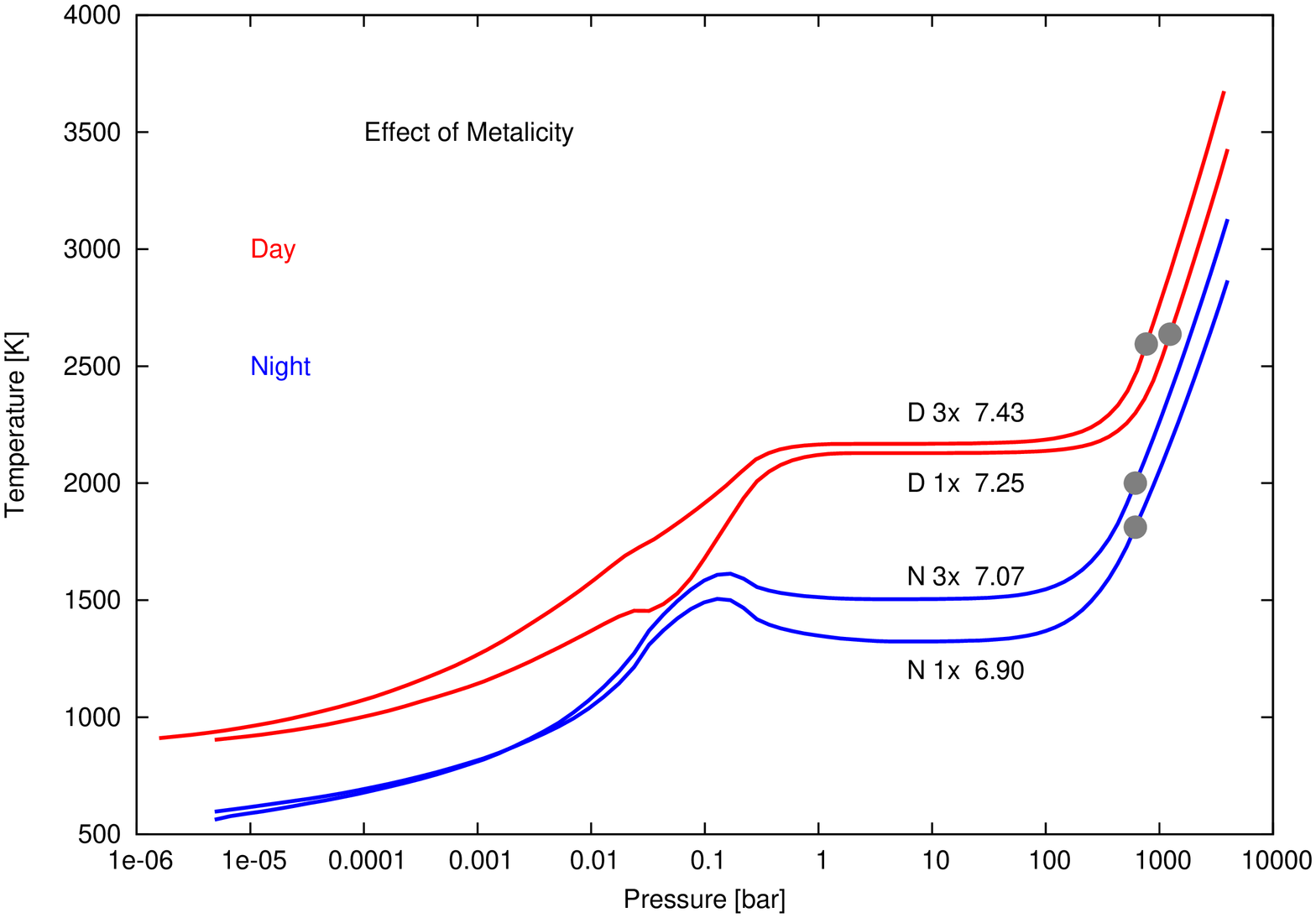}
\includegraphics[angle=-90,width=8.5cm]{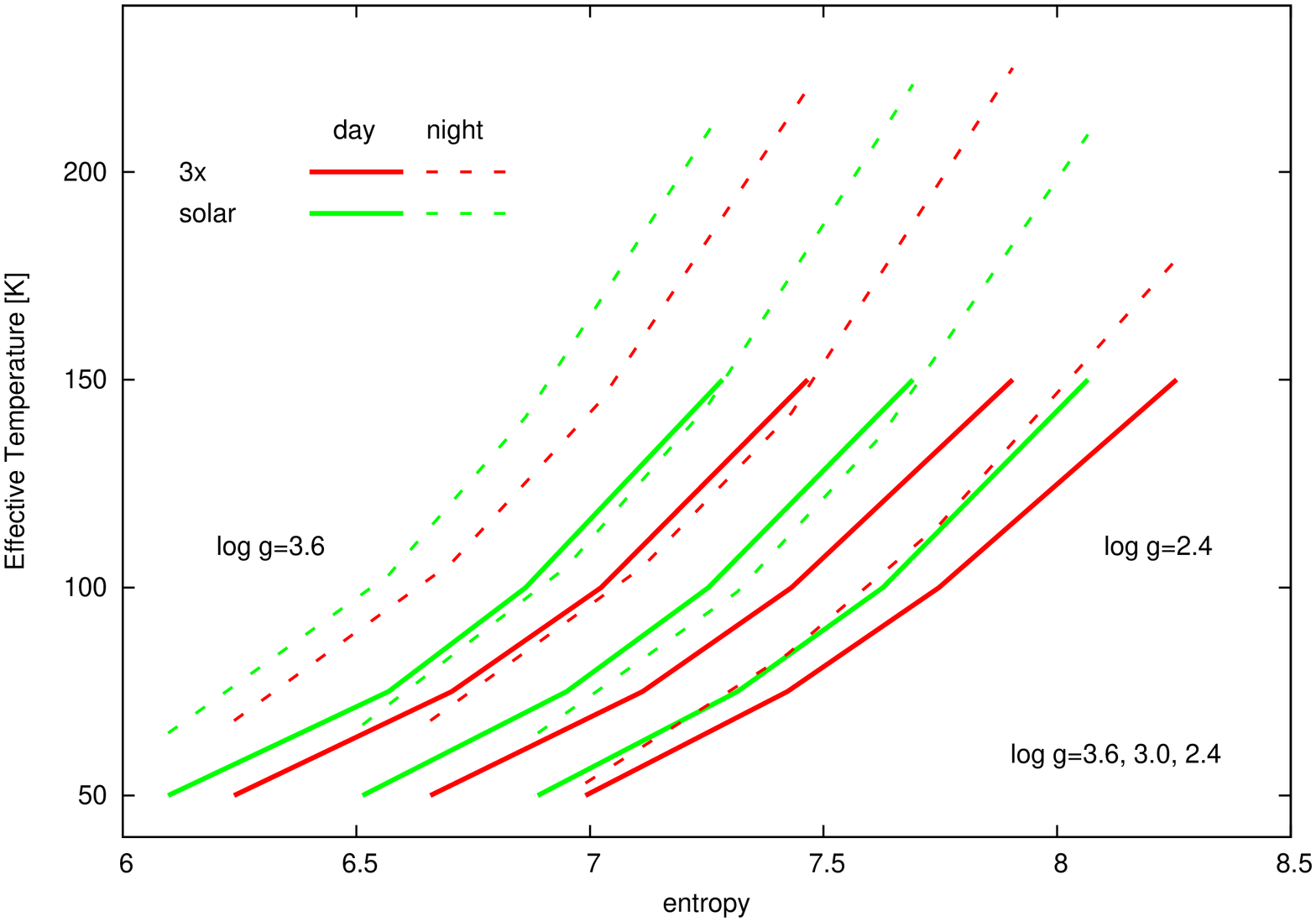} 
}
\caption{
{\bf Left: The effect of varying the metallicity on the $T$-$P$ profile of 
the atmosphere.}
The curves are labeled by the value of the metallicity in the atmosphere 
(1 or 3x solar) and the core entropy, respectively.
Notice that entropy on both day and night sides increases with increasing
metallicity.
{\bf Right: The effect of varying the metallicity of the planetary
atmosphere on the day-night side cooling.} Models are 
calculated for two metallicities: solar and 3x solar metallicity.
The cooling from both the day and the night side decreases with 
the metallicity. Consequently, the total heat loss decreases with 
increasing metallicity.
}
\label{met1tp}
\end{figure*}

In this section we investigate the effect of enhanced metallicity
(and, therefore, opacity) on the cooling of the planetary interior.
Burrows et al. (\cite{bhb07}) have already demonstrated that 
an enhanced opacity in the planetary atmosphere due to higher metallicity 
reduces the escape of internal heat.  They considered $P_{n}=0.5$, and 
assumed that the night side cools at the same rate as the day side.

To study the effects of metallicity on the day- and night-side cooling 
rates we calculate a set of day and night atmosphere models for two 
different metallicities. One corresponds to solar metallicity, and 
the other is enhanced by 0.5 dex, i.e. a factor of about 3.   
We assume $P_{n}=0.3$; the incoming irradiation energy is removed at 
depths between 0.03 and 0.3 bar on the day side, and is deposited
on the night side in the same pressure region.

Figure \ref{met1tp} (left) illustrates the effect of metallicity
on the $T$-$P$ profiles, for a model with $T_{\rm eff}=100$ K and 
$\log_{10} g=3$.
The day-side temperature plateau level is not very sensitive to 
metallicity. However, for the lower metallicity the plateau
extends much deeper into the interior, where pressure and density are higher.
This is easily understood by noticing that lower metallicity results in
a lower opacity; consequently the atmosphere is more transparent and 
the effect of irradiation is felt at deeper layers.
Pressure at the top of the convection zone is lower for higher 
metallicity. Consequently, given the relation from Figure \ref{ent},
entropy is higher (at a particular gravity and 
intrinsic effective temperature) for models with higher metallicity.
This means that one has to lower the intrinsic effective temperature
(i.e., lower the cooling) to fit the same entropy.
The night side behaves analogously. Consequently, the total core cooling 
rate is also reduced for higher metallicity models, as expected. 
Figure \ref{met1tp} (right) depicts this cooling as a function 
of entropy for several gravities. This demonstrates that the above mentioned 
behavior holds for the whole grid of models with a range of entropies 
and gravities. Figure \ref{dd2} (right) depicts the total combined day+night 
cooling for the whole grid of models.

In conclusion, increased metallicity (opacity) acts as a blanket that 
inhibits heat loss, and, thus, keeps the planet warm for a longer period 
of time.

\section{The effect of the $P_{n}$ parameter}
\label{s4}

The $P_{n}$ parameter is an empirical parameter that specifies 
the portion of irradiated energy received by the day side that is
transported to the night side (Burrows, Sudarsky, \& Hubeny \cite{bsh06}),
and thus measures the efficiency of day-night heat transfer. 
$P_{n}=0$ means that there is no
heat flow from the day to the night side, while $P_{n}=0.5$ means  
that half of the energy received by the day side is carried 
to the night side and re-radiated from there.

The value of $P_{n}$ can be constrained by secondary eclipse observations.
It turned out that its value may differ from planet to planet.
Harrington et al. (\cite{hhl06}) favor 
a small amount of the heat redistribution for $\upsilon$ And. 
Cowan, Agol \& Charbonneau (\cite{cac07}) estimated that
$P_{n}\leq0.30\,(1 \sigma)$ for HD~179949b, while 
$P_{n}\geq0.32\,(1\sigma)$ for HD~209458b. 
Burrows, Budaj \& Hubeny (\cite{bbh08}) estimated heat redistribution
efficiencies for six transiting hot Jupiters.
Cowan \& Agol (\cite{ca11}) 
studied Bond albedos and heat redistribution of a sample
of 24 exoplanets. Smith et al. (\cite{s11}) found low heat redistribution 
for WASP-33b.
Anderson et al. (\cite{anderson10}), Gibson et al. (\cite{gibson10}) and 
Budaj (\cite{budaj11}) found low heat redistribution for WASP-19b.

To investigate the effect of day-night heat transfer
on day and night side core cooling we calculate a set of day and night
side atmosphere models for different values (0.1, 0.2, 0.3, 0.4, 0.5) 
of the $P_{n}$ parameter. 
We assume that the incoming irradiation energy is removed
between 0.03 and 0.3 bar on the day side, and is deposited
on the night side in the same pressure region.

\begin{figure*}
\centerline{
\includegraphics[angle=-90,width=8.5cm]{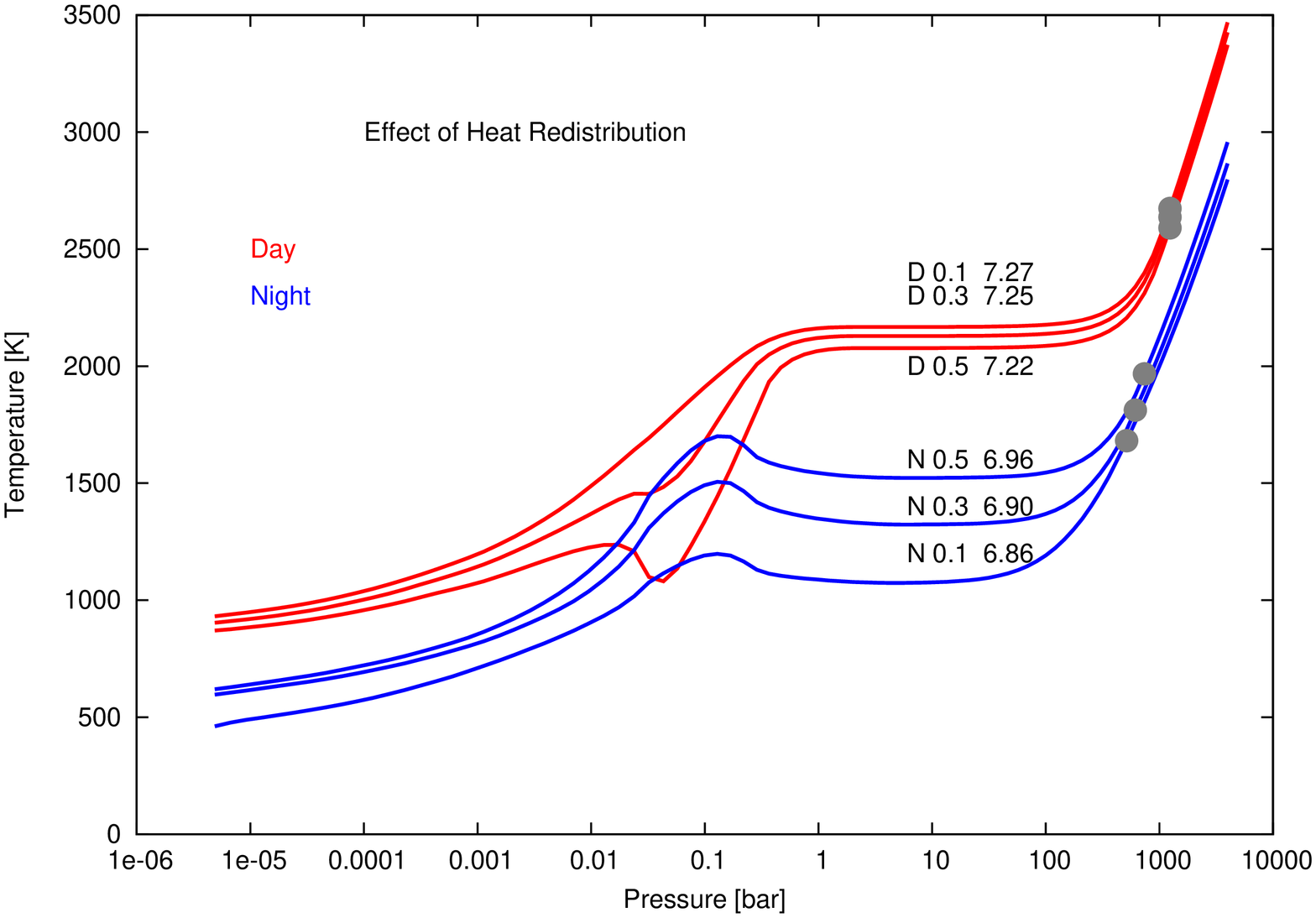}
\includegraphics[angle=-90,width=8.5cm]{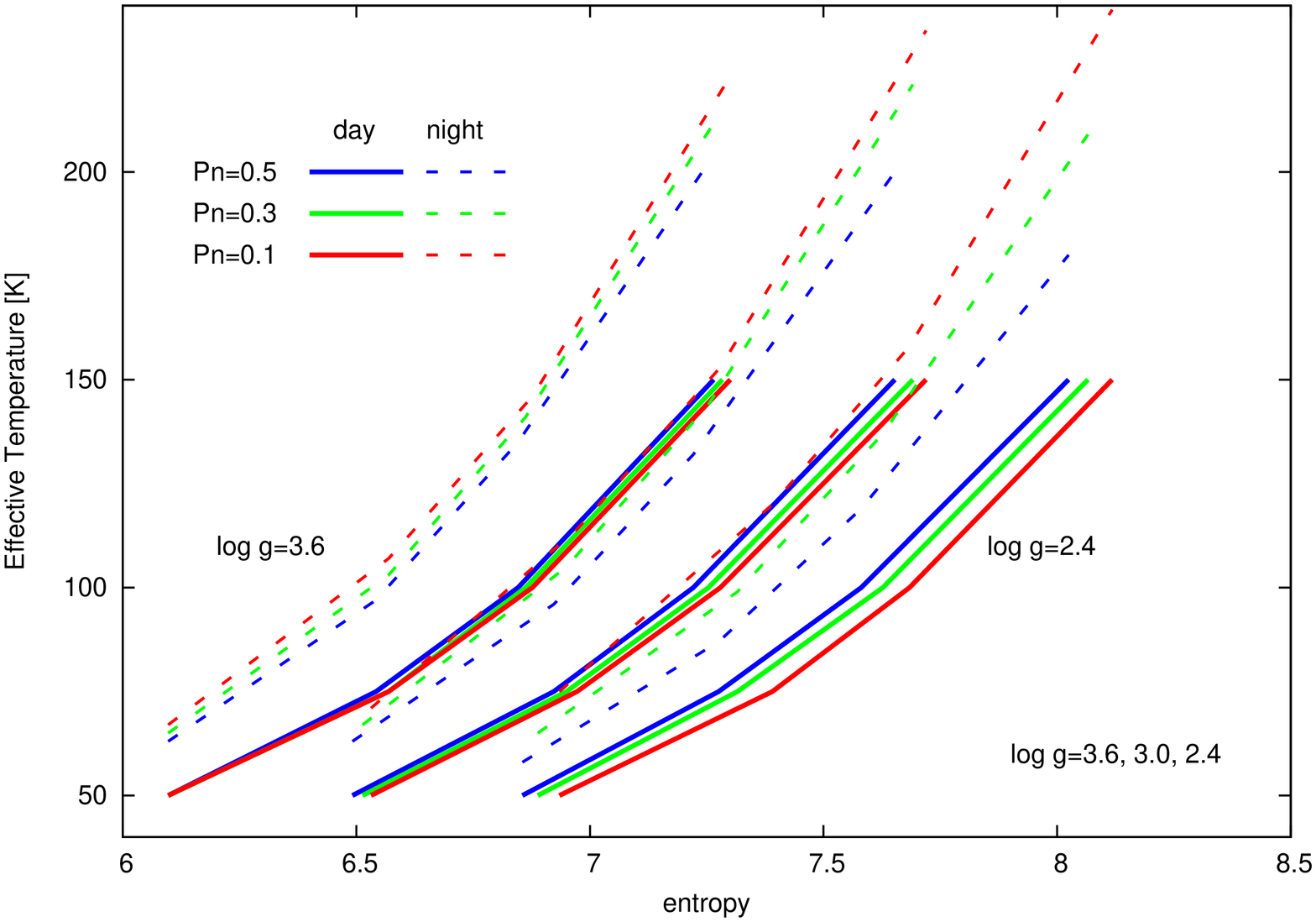}
}
\caption{
{\bf Left: The effect of the $P_{n}$ parameter (day-night 
heat transfer efficiency) on the $T$-$P$ profile of the atmosphere.}
The curves are labeled with the values of the $P_{n}$ parameter and 
the core entropy, respectively.
Notice that the entropy on the day side decreases with increased 
efficiency of heat
redistribution. Entropy on the night side behaves in the opposite way
and increases with increasing heat redistribution, or $P_{n}$ value.
{\bf Right: The effect of the $P_{n}$ parameter
on the day - night side cooling}. The models are
calculated for three different values of $P_{n}=0.1, 0.3, 0.5$.
Cooling from the night side decreases with the efficiency of
heat redistribution ($P_{n}$). Cooling from
the day side behaves in the opposite way and increases with increasing
$P_{n}$. However, the night side is more sensitive to the $P_{n}$ parameter,
so it governs total core heat loss, which decreases with increasing
$P_{n}$. The difference between night and the day
side cooling is largest for the smallest values of $P_{n}$.
}
\label{Pn1tp}
\end{figure*}

Figure \ref{Pn1tp} (left) illustrates the $T$-$P$ profiles
for a model with $T_{\rm eff}=100$ K and $\log_{10} g=3$.
On the day side the temperatures on the plateau and at the top of 
the convection zone
decrease with increasing $P_{n}$. Subsequently (see Figure \ref{ent}),
the entropy in the convection zone on the day side decreases 
with increasing $P_{n}$, all else being equal.
This means that cooling (that is, $T_{d}$) must increase for higher $P_{n}$ 
values (at the same entropy and gravity), thus, day side heat loss is higher.
On the other hand, the plateau temperature and the entropy on the night 
side increase with increasing $P_{n}$. 
This means that for higher $P_{n}$ (at the same
entropy and gravity) $T_{n}$ must be lower, and, thus, the losses from   
the night side are lower.
Figure \ref{Pn1tp} (right) depicts cooling as a function of entropy for
several gravities, and shows that the same trend is valid for the whole 
grid of models.
Notice that the difference between night and day cooling $T_{n}-T_{d}$ 
is higher for low heat redistribution, and decreases with more efficient 
day-night heat transport.

\begin{figure*}
\centerline{
\includegraphics[angle=-90,width=10.cm]{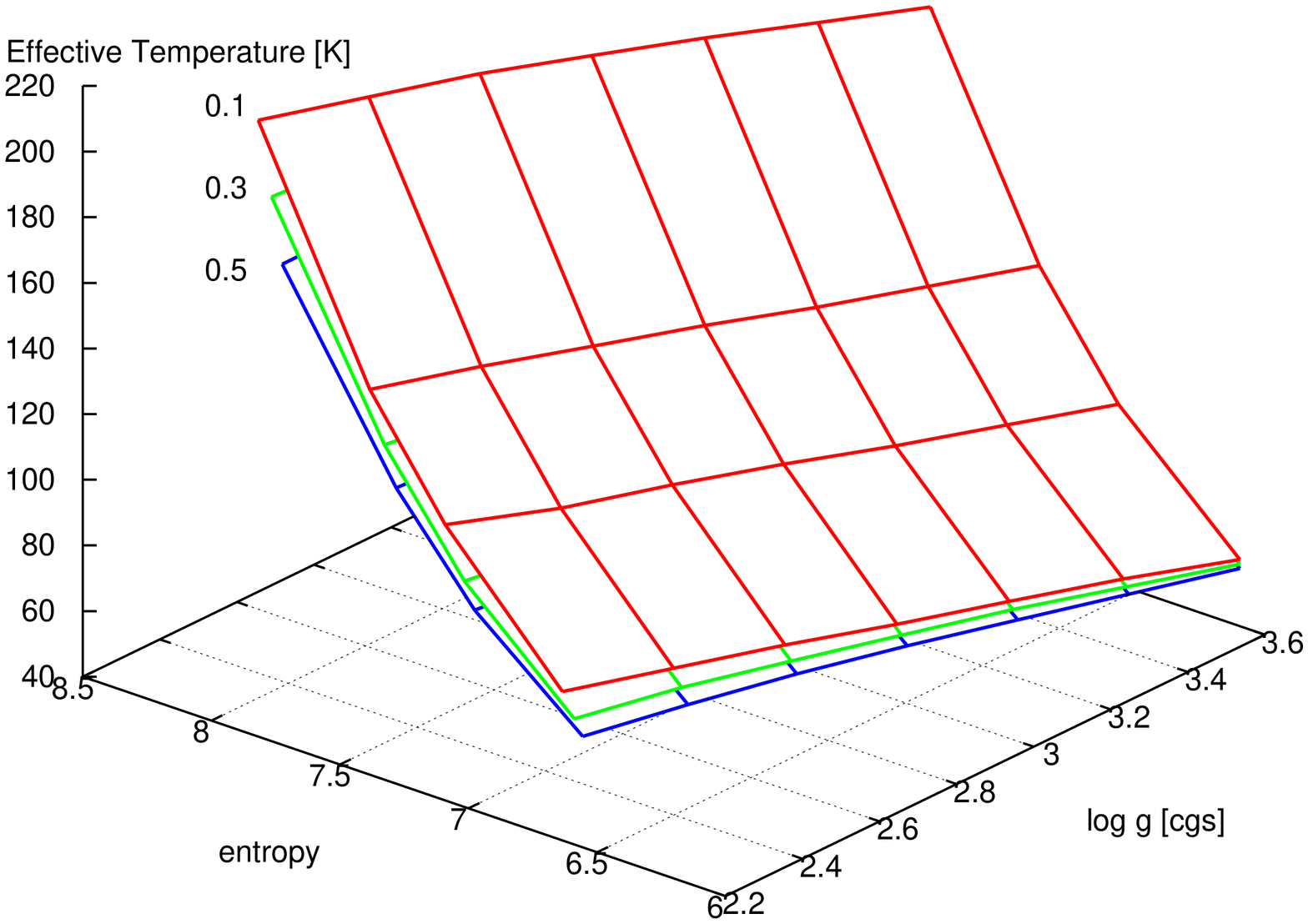}
\hspace{-10mm}
\includegraphics[angle=-90,width=10.cm]{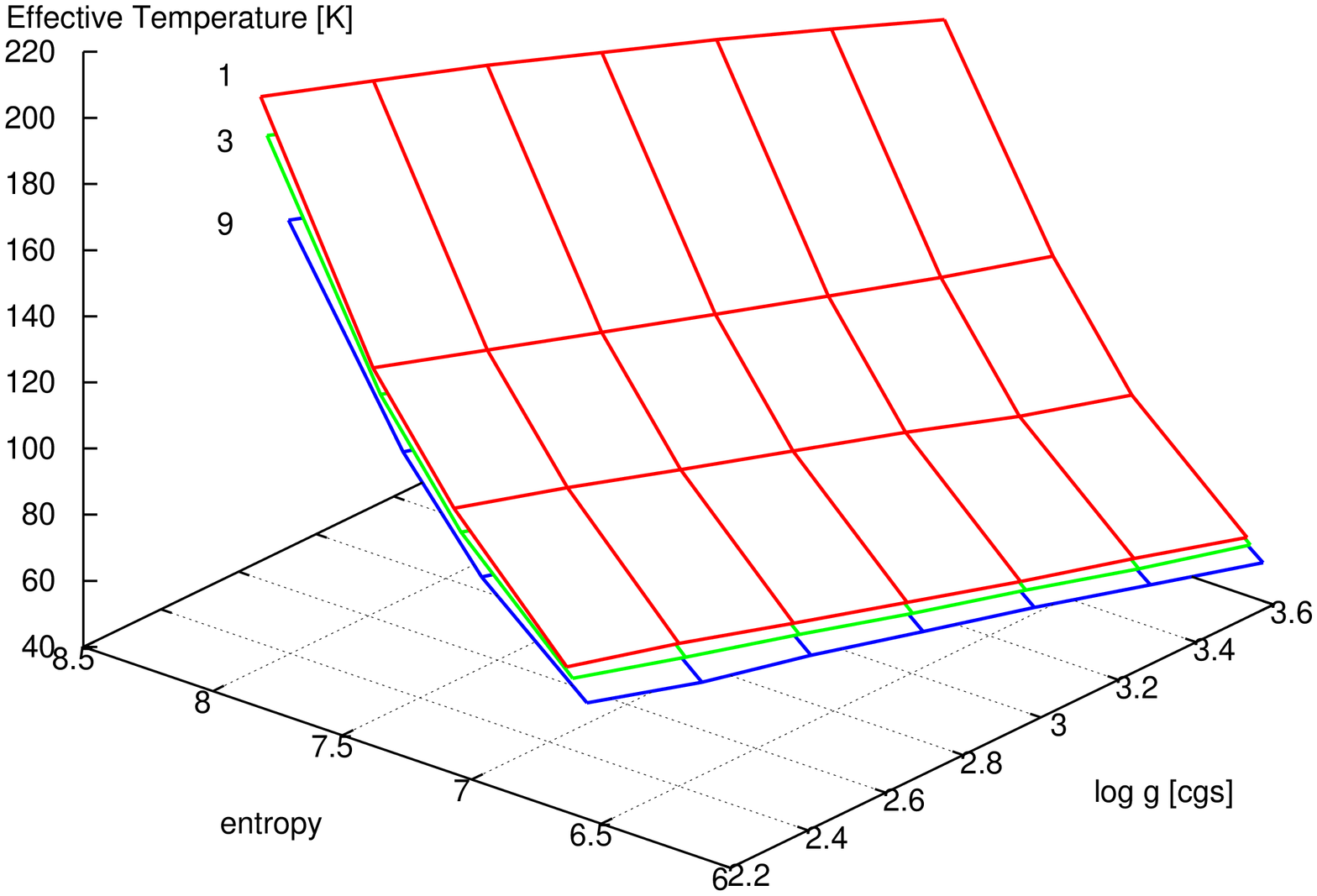}
}
\caption{
The combined night and day side cooling
as a function of the surface gravity and core entropy.
It is expressed as the intrinsic effective temperature
$T_{\rm eff}=(T_{d}^{4}/2+T_{n}^{4}/2)^{0.25}$ in $K$.
{\bf Left: The effect of the $P_{n}$ parameter.} The models are
calculated for three different values of $P_{n}=0.1, 0.3, 0.5$.
The total core heat loss decreases with increasing $P_{n}$. 
{\bf Right: The effect of the depth of the heat redistribution.} The models are
calculated for three different values of the day side Rosseland optical 
depth of the bottom of the heat redistribution region $\tau_{\rm ross}=1, 3, 9$.
Total heat loss decreases with increasing optical depth.
See the text for a more detailed discussion.
}
\label{pn2}
\end{figure*}

In this case, the day and the night sides behave in opposite ways.
However, the night side is more sensitive to heat transfer
and the total (day plus night side) heat loss decreases
with increasing heat transfer.
This is shown in the Figure \ref{pn2} (left).
This behavior can be understood in the following way.
On the night side, a higher $P_{n}$ value means more energy deposited
at a certain depth. This has a similar effect on the $T$-$P$ profiles 
as stronger irradiation and the outcome is the same -- a reduced heat loss. 
The day side with an energy removal behaves in the opposite way.

\section{The effect of extra opacity in the upper atmosphere
of the day side of the planet}
\label{s5}

\begin{figure*}
\centerline{
\includegraphics[angle=-90,width=8.5cm]{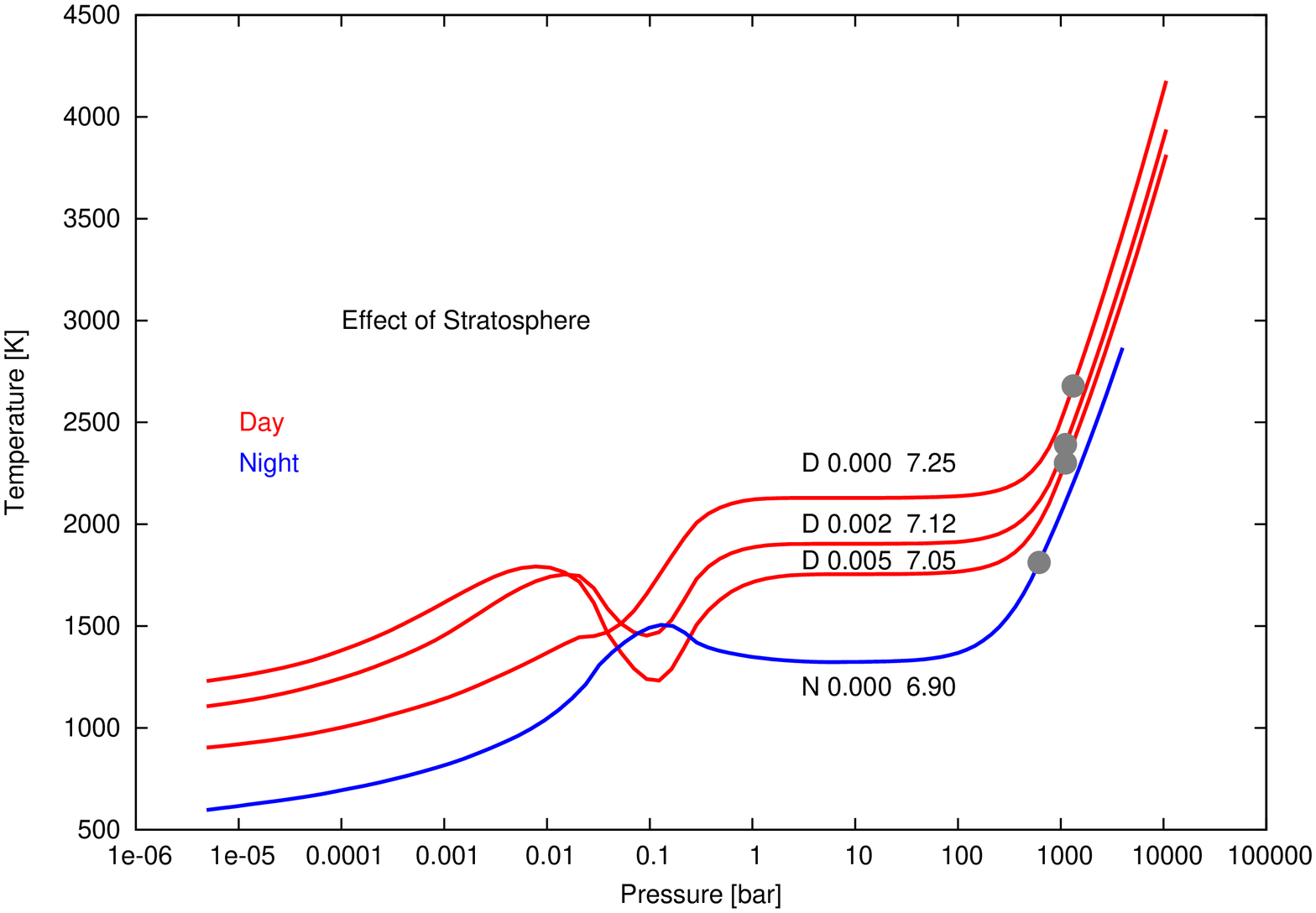}
\includegraphics[angle=-90,width=8.5cm]{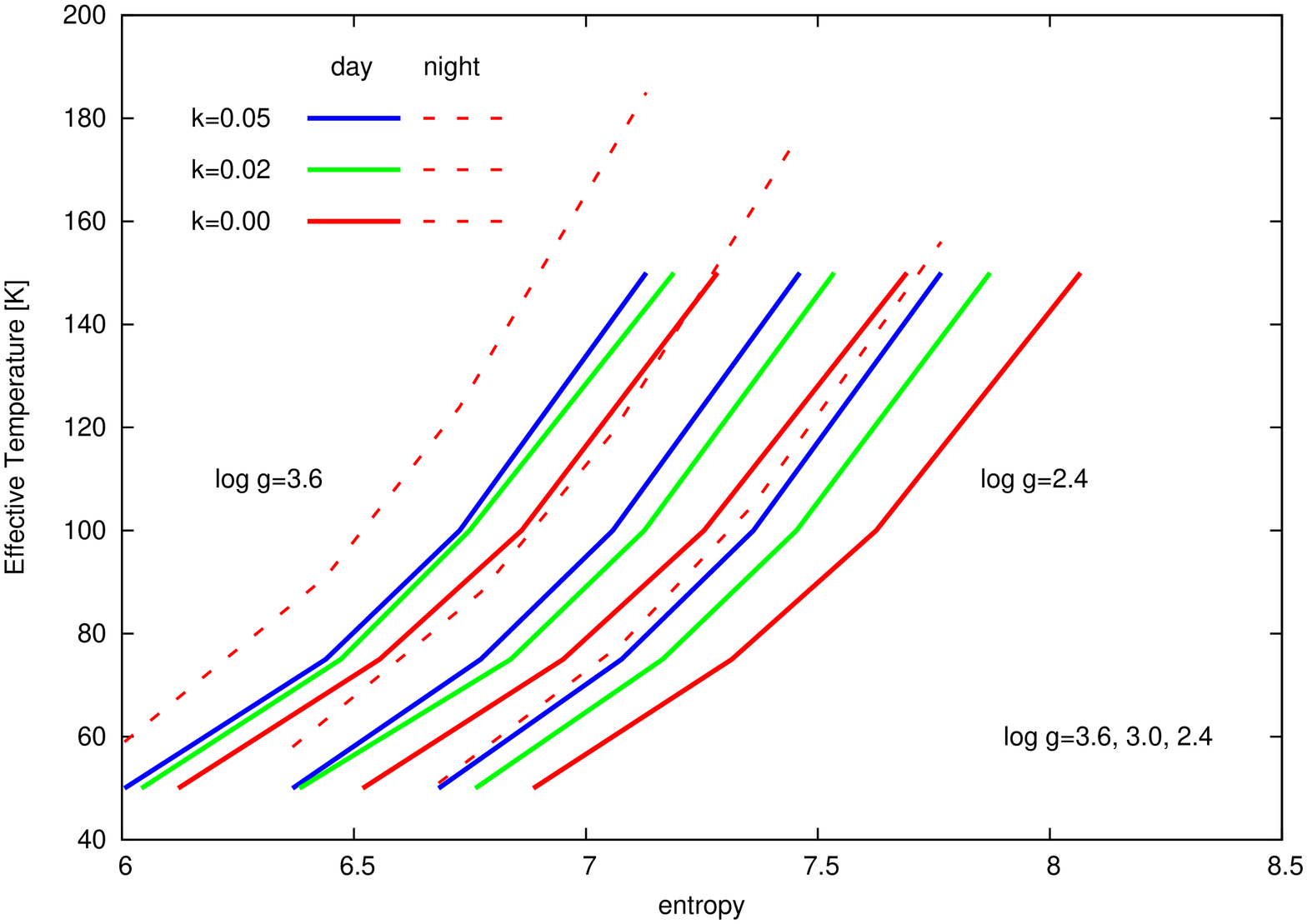}
}
\caption{
{\bf Left: The effect of a stratospheric absorber on the $T$-$P$ profile 
of the day-side atmosphere.}
The curves are labeled with the values of the extra opacity, 
$\kappa=0., 0.02, 0.05$~cm$^{2}$g$^{-1}$, and the core entropy, respectively.
Notice that the entropy on the day side decreases with increasing the extra opacity.
There is no stratosphere on the night side, so only one model is plotted.
{\bf Right: The effect of the extra opacity in the upper atmosphere
of the day side on the day - night side cooling.} The models are
calculated for three different values of the extra opacity,
$\kappa=0., 0.02, 0.05$~cm$^{2}$g$^{-1}$.
Cooling from the day side increases with the value of the extra
opacity. Cooling from the night side core is the same since only the day
side exhibits a stratosphere.
The difference between the night and the day
side cooling is largest for lower values of the extra opacity, and 
the total core heat loss increases with increasing extra opacity.
}
\label{kappa1tp}
\end{figure*}

Hubeny, Burrows \& Sudarsky (\cite{hbs03}) first pointed out that 
highly irradiated cold atmospheres with an efficient opacity source
in the optical could exist in two distinct configurations;
one with a temperature monotonically decreasing outward, and one 
exhibiting a temperature increase toward a surface. They also showed that
a strong temperature inversion at the top of the atmosphere
would have a significant effect on the emergent spectrum.
Early spectrophotometric observations of the secondary eclipses indeed
gave a strong indication of a
temperature inversion on 
the day side of some close-in planets (Burrows et al. \cite{bhbk07};
Fortney et al. \cite{flm08}).
Some attributed the opacity that leads to such a temperature
inversion to TiO and VO. However, there is uncertainty as to whether TiO and
VO can indeed be supported in the upper atmosphere, that is whether
one can find an efficient mechanism that would prevent TiO and VO from
settling downward in the atmosphere to colder regions where they would be
rained out (the so-called cold trap effect -- Spiegel et al. \cite{ssb09}).
Therefore, given the uncertainty about the actual opacity source that
is responsible for the creation of the ``stratosphere'', we use
the term ``extra'' opacity and parametrized its value and
the wavelength range in which it operates (Burrows, Budaj, \& Hubeny \cite{bbh08}).

Here, we investigate how such an extra opacity
on the day side can affect the day-night side core cooling of the planet.
We calculate a grid of day- and night-side atmosphere models
with an extra absorber with various opacities,
$\kappa=0., 0.02, 0.05$ g\,cm\,$^{-2}$ corresponding to optical depths
of 0, 0.5, 1.4.
The extra absorption is gray, operating only in the optical region  
for wavelengths between 0.3 and 1 micron, and placed only in the 
upper atmosphere of the day side. 
The efficiency of heat redistribution is set to $P_{n}=0.3$, and
we assume that the incoming stellar energy is removed at the
depth region between 0.03 and 0.3 bar at the day side, and is deposited
on the night side over the same pressure region.

Figure \ref{kappa1tp} (left) shows the effect of the  
absorber on the $T$-$P$ profile
for a model with $T_{\rm eff}=100$ K and $\log_{10} g=3$.
The day-side plateau and temperature at the top of the convection zone
decrease with increasing extra opacity. Consequently (see Figure \ref{ent}), 
the core entropy decreases with increasing extra opacity, 
and, thus, the heat loss from the day side must be higher if the entropy
is to be constant.
At first glance this might seem strange because 
it is just the opposite of the effect of increasing the metallicity 
(see \S\ref{s3a}).
The explanation is that a higher metallicity leads to a higher opacity
throughout the whole atmosphere, and, thus, it acts as a blanket   
that reduces the heat loss from the day side.
In contrast, a higher extra stratospheric opacity works differently. 
It acts more like a ``shield'' that prevents the energy flux from 
the star from penetrating deeper into the planet atmosphere
because part of the irradiation energy 
is now deposited high in the atmosphere.

\begin{figure}
\centerline{
\includegraphics[angle=-90,width=10.cm]{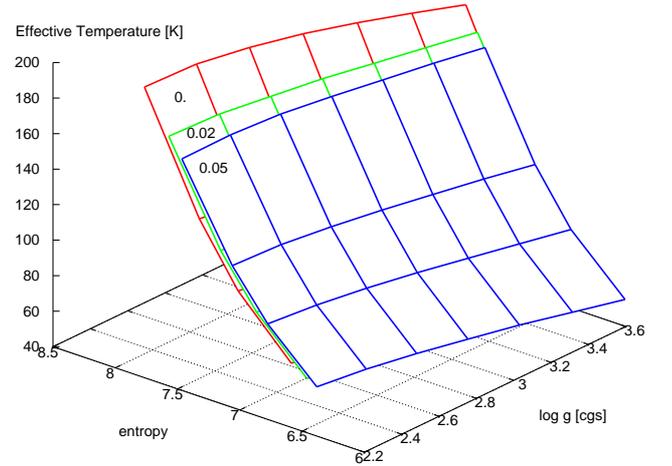}
}
\caption{
The combined night and day side cooling
as a function of the surface gravity and core entropy.
It is expressed as the intrinsic effective temperature
$T_{\rm eff}=(T_{d}^{4}/2+T_{n}^{4}/2)^{0.25}$ in $K$.
{\bf The effect of the stratosphere.} The models are
calculated for three different values of extra opacity 
$\kappa=0., 0.02, 0.05$.
The total core heat loss increases with increasing stratospheric opacity. 
}
\label{kappa2}
\end{figure}

This ``shield'' becomes hotter with increasing extra opacity,
but the layers below become cooler since they are optically thin at other
wavelengths. Speaking figuratively, the overall effect resembles a breeze 
that carries heat out and creates a cool and comfortable area under 
a hot sunshade on a sunny beach. 
Thus, the behaviour of the temperature plateau is analogous to the situation 
studied in Sec.\ref{s3} when the planet was farther away from the star, which 
results in higher day side cooling for higher extra opacity.
Figure \ref{kappa1tp} (right) illustrates the cooling effect as a function
of the entropy for several gravities and shows that this result holds for 
the whole grid of models.
Figure \ref{kappa2} shows the total combined day+night cooling which 
increases with increasing extra opacity.

\section{The effect of the depth of day-night heat transport region}
\label{s6}

In the above mentioned calculations, we kept the region of day-night
heat redistribution constant in pressure and equal for both the day and    
night sides for the whole grid of models, while either the planet-star 
separation, metallicity, $P_{n}$, or extra opacity were varied.
However, the optical depth at a certain pressure level is a strong function of
gravity. The actual day-night heat redistribution may take place
within a certain range of optical depths, rather than in a constant pressure
range.
This is why we calculate a set of models where, on the day side, we adjust
the region of the day-night heat redistribution so that   
the Rosseland optical depth of the region is constant.
We study three cases, and place the energy sink on the day side at
$\tau_{\rm Ross}= 0.06-1,\ 0.2-3$, and $0.6-9$.
On the night side, the energy is deposited at the same pressures
(corresponding to the above mentioned optical depths) as the energy is
being removed on the day side. In all models, we assume $P_{n}=0.3$.

\begin{figure*}
\centerline{
\includegraphics[angle=-90,width=8.5cm]{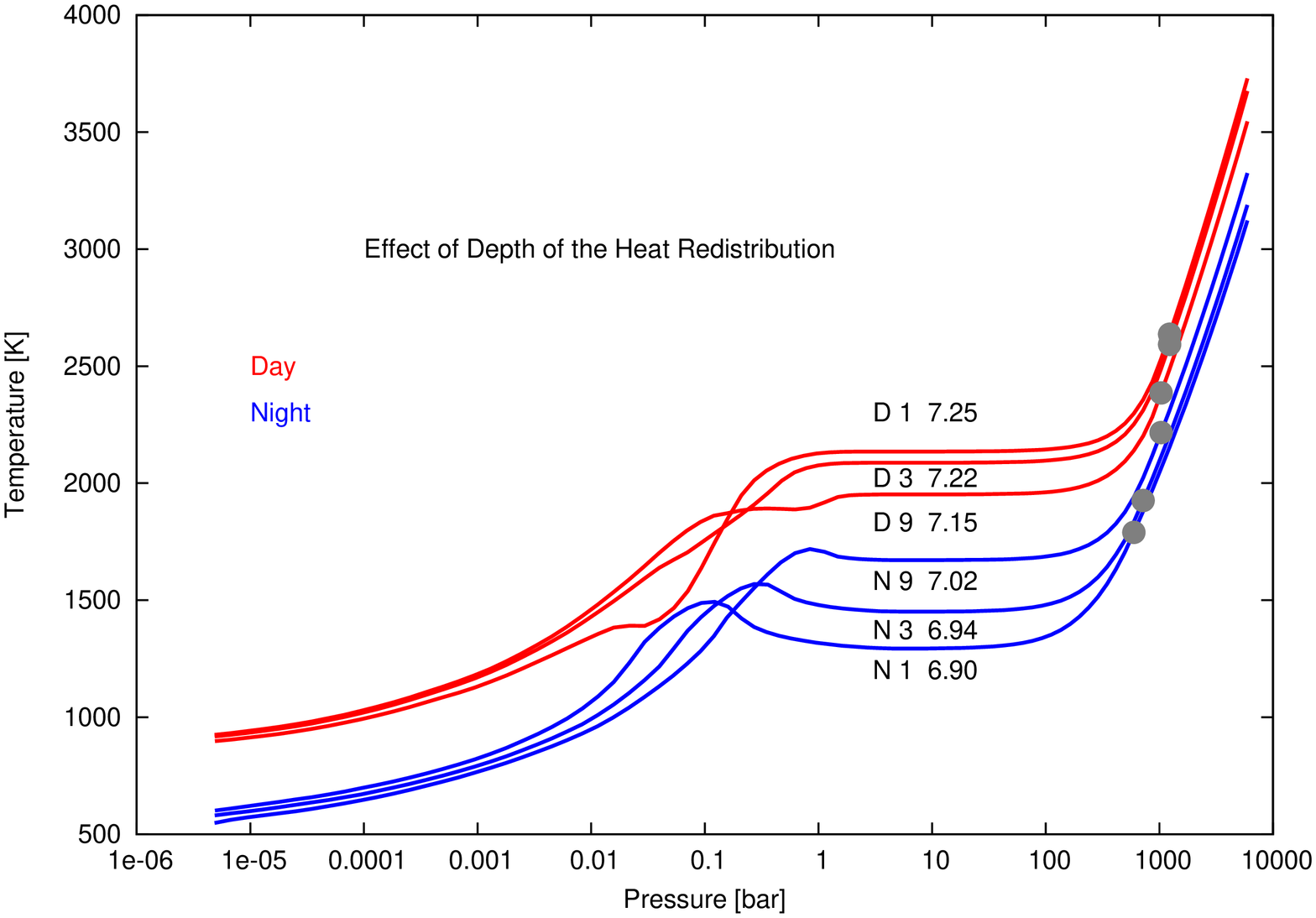}
\includegraphics[angle=-90,width=8.5cm]{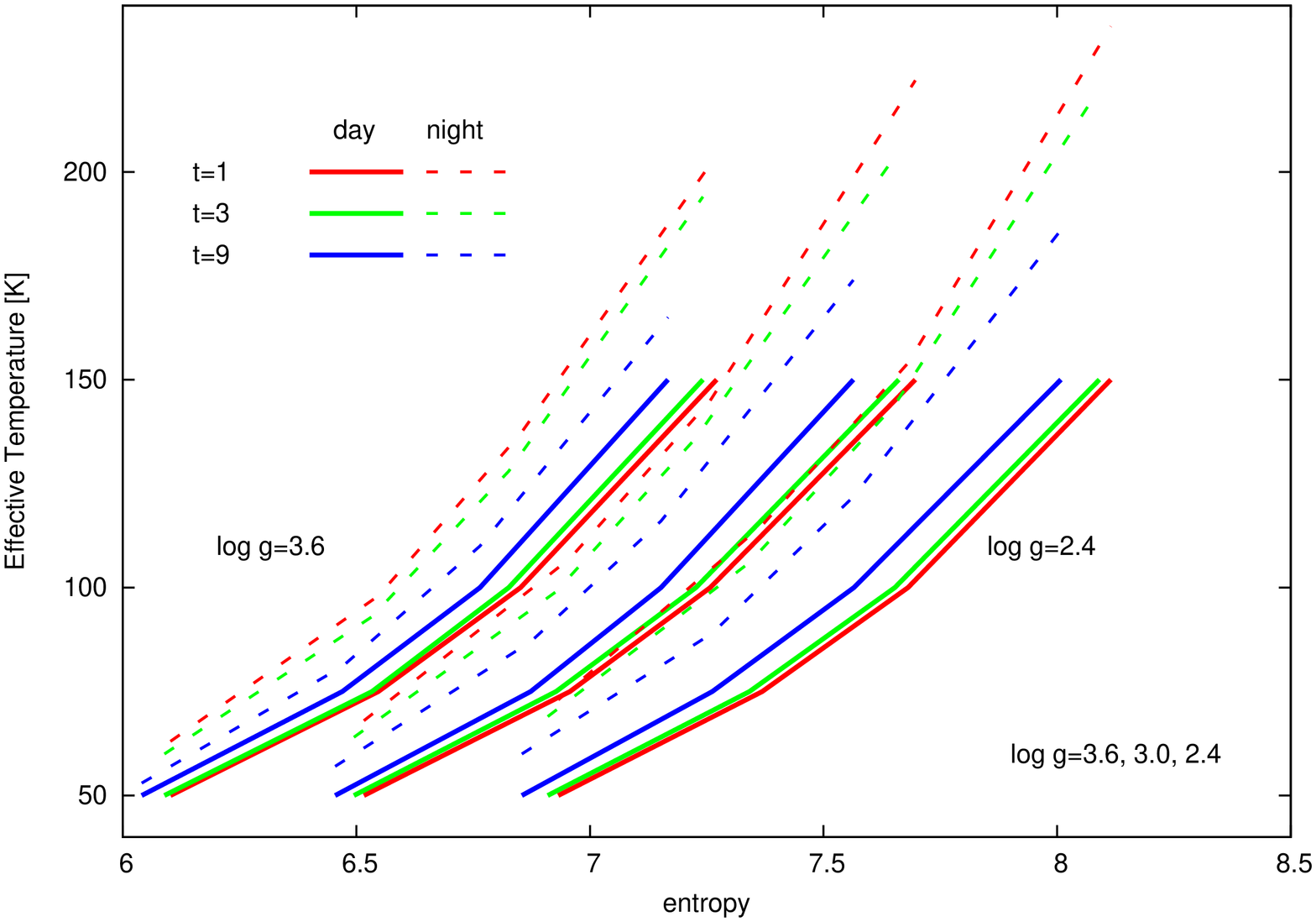}
}
\caption{
{\bf Left: The effect of the depth of the day-night heat transport region on 
the $T$-$P$ profile of the atmosphere.}
The curves are labeled with the values of the Rosseland optical depth 
of the bottom of the day-night heat transfer region,
and the core entropy, respectively.
Notice that the entropy on the day side decreases with increasing optical depth.
The entropy on the night side behaves in the opposite way, and increases
with the optical depth of the heat redistribution region.
{\bf Right: The effect of the depth of the day-night heat transport
on the day - night side cooling.}
The day-side (solid) and the night-side (dashed) cooling
of the planet for three different values of the day-side Rosseland optical
depth at the bottom  of the heat redistribution region $\tau_{\rm ross}=1, 3, 9$.
On the day side, the heat loss increases with increasing optical depth.
On the night side, the heat loss decreases with increasing optical depth
and this trend tends to dominate the opposite trend on the day side.
As a result the night-to-day difference in heat loss is lower for larger optical depth
and the total heat loss decreases with increasing the optical depth.
}
\label{tau1tp}
\end{figure*}

Figure \ref{tau1tp} (left) illustrates the effect of varying the depth 
of the heat redistribution region on the $T$-$P$ profiles.
On the day side, the plateau and temperature at the top of the
convection zone decrease with increasing optical depth.
Consequently, (see Figure \ref{ent}) the core entropy decreases 
with increasing optical depth. This means that heat loss is higher 
at the same entropy and gravity. On the night side the plateau temperature 
and entropy increase with 
increasing optical depth, which means that the heat loss is lower (at 
the same entropy and gravity) for higher optical depth. 
This is shown in Figure \ref{tau1tp} (right), where we plot the cooling 
rate as a function of entropy for several gravities.
The night side is more sensitive to the optical depth, and the total heat loss 
decreases with increasing depth of the day-night heat redistribution
region (see Figure \ref{pn2}-right).
This behavior is very similar to the dependence of the heat loss 
on the $P_{n}$ parameter.
This is not surprising, since the effect of increasing the depth of 
heat redistribution on the model atmosphere is analogous
to the effect of an increase of the heat redistribution efficiency.
Moreover, in reality these two effects are probably correlated, in the sense 
that if heat redistribution takes place at deeper layers it is likely to be
more efficient, and, thus, to be described by a higher $P_{n}$ value, which 
has the same effect on day-night side cooling. Consequently, their 
mutual correlation could make the effect stronger.

In all the above calculations we assumed that
the heat redistribution region on the night side corresponds
to the same pressure range as on the day side.
Based on these calculations one can
speculate what will happen if this is not the case.
The 3D models of Dobbs-Dixon \& Lin (\cite{dl08}) 
and Showman et al. (\cite{scf08}) indicate that
this problem is very complex. The eastward and westward flows can
circumnavigate the planet at different latitudes, and one can sink
below the other depending on the cooling they experience.
If energy is deposited on the night side at lower pressures than those
at which energy is being removed on the day side then we would observe  
an enhanced amount of core cooling from the night side.
On the other hand, if energy is deposited at higher pressures
on the night side than energy is removed from on the day side, 
night-side cooling could be significantly reduced.

\section{Conclusions}

We have studied the effects of several mechanisms on day- and night-side
core cooling of strongly irradiated planets. 
The main conclusions are:
\begin{itemize}

\item
We have demonstrated that the cooling of a strongly irradiated giant planet
is different on the day and the night sides.

\item
An increased amount of planet irradiation leads to reduced 
day and night side cooling, and thus to reduced total core heat loss.

\item
An increased metallicity throughout the atmosphere leads to
reduced day and night side cooling, and thus to reduced total core heat loss.

\item
An increased efficiency of the day-night heat redistribution reduces
heat loss from the night side, while increasing heat loss    
from the day side. The night side is more sensitive to this effect, 
and as a result the total heat loss is reduced.

\item
A possible extra opacity in the stratosphere on the day side of the planet
acts as a sunshade that partly blocks the irradiation from the star.
Consequently, it increases the day side cooling as well as the total heat
loss.

\item
If day-night heat redistribution takes place at larger optical depths
it increases day side core cooling, while night side core cooling is
suppressed. The total heat loss is also reduced.

\item
Night side cooling is generally more efficient than the day side
cooling. However, this does not rule out the possibility that 
there is a combination of parameters where the opposite conclusion 
applies. 
\end{itemize}

These effects may influence the evolution of strongly irradiated 
substellar-mass objects. If a particular effect results in enhanced
core cooling this leads to faster shrinking and lower radii and vice versa.
These effects will also compete with gravity darkening and bolometric albedos 
in interacting binaries 
(von Zeipel \cite{zeipel24}; Lucy \cite{lucy67}; Rucinski \cite{rucinski69};
Vaz \& Norlund \cite{vn85}; Claret \cite{claret98}), especially for strongly 
irradiated cold components. 
As a result one might observe a sort of ``day side darkening'' or 
``night side brightening'' in the intrinsic flux coming from the interior
of an irradiated object.

\begin{acknowledgements}
The work of JB has been supported by the VEGA grants of the Slovak Academy of
Sciences Nos. 2/0074/09, 2/0078/10, 2/0094/11. IH wishes to thank the
Slovak Academic Information Agency SAIA for a support during his stay at 
the Astronomical Institute of SAV where the work was completed.
We also thank Prof. Saumon and an anonymous referee for their comments 
and suggestions on the manuscript.

\end{acknowledgements}

\end{document}